\documentclass[aps,prl,twocolumn,showpacs]{revtex4}
\usepackage[english]{babel}
\usepackage{dcolumn}
\usepackage{graphicx}
\usepackage{amsmath}
\usepackage{longtable}
\usepackage{bigstrut}

\begin{document}

%\preprint{}

\title{A molecular fountain}

\author{Cunfeng Cheng}
\author{Aernout P.P. van der Poel}
\author{Paul Jansen}
\thanks{Currently at: Laboratory of Physical Chemistry, ETH Zurich, CH-8093 Zurich, Switzerland.}
\author{Marina Quintero-P\'{e}rez}
\thanks{Currently at: Netherlands Organisation for Applied Scientific Research (TNO), Stieltjesweg 1, 2628 CK Delft, The Netherlands.}
\author{Thomas E. Wall}
\thanks{Currently at: Centre for Cold Matter, Blackett Laboratory, Imperial College London, Prince Consort Road, London SW7 2AZ, UK.}
\author{Wim Ubachs}
\author{Hendrick L. Bethlem}
\thanks{Electronic address: H.L.Bethlem@vu.nl}
\affiliation{LaserLaB, Department of Physics and Astronomy, VU University Amsterdam,
De Boelelaan 1081, 1081 HV Amsterdam, The Netherlands} 
%\thanks{}


\date{\today}

\begin{abstract}
The resolution of any spectroscopic or interferometric experiment is ultimately limited by the total time a particle is interrogated. We here demonstrate the first molecular fountain, a development which permits hitherto unattainably long interrogation times with molecules. In our experiments, ammonia molecules are decelerated and cooled using electric fields, launched upwards with a velocity between 1.4 and 1.9\,m/s and observed as they fall back under gravity. A combination of quadrupole lenses and bunching elements is used to shape the beam such that it has a large position spread and a small velocity spread (corresponding to a transverse temperature of  $<$10\,$\mu$K and a longitudinal temperature of  $<$1\,$\mu$K) when the molecules are in free fall, while being strongly focused at the detection region. The molecules are in free fall for up to 266\,milliseconds, making it possible to perform sub-Hz measurements in molecular systems and paving the way for stringent tests of fundamental physics theories.  
\end{abstract}

\maketitle

The ability to control atoms using lasers~\cite{Chu:RMP1998, Cohen-Tannoudji:RMP1998, Phillips:RMP1998} has resulted, amongst many other spectacular achievements, in the demonstration of an atomic fountain~\cite{Kasevich:PRL1989}. In such a fountain, laser cooled atoms are gently pushed upwards on a vertical trajectory and left to fall back under gravity. The long free fall times permitted by a fountain allow very precise spectroscopy to be performed. In an atomic fountain clock~\cite{Clairon:EPL1991}, the atoms pass a microwave cavity twice -- as they fly up and as they fall back down. The effective interrogation time in such a Ramsey-type measurement scheme includes the entire flight time between the two traversals through the driving field. This interrogation time is typically up to one second. Nowadays, about a dozen fountain clocks are operated at various metrological institutes around the world to realize the SI unit of time, the second, with an accuracy better than 10$^{-15}$~\cite{Wynands:Met2005,Guena:IEEE2012, Heavner:Met2014}. Atomic fountains are also used in matter-wave interferometry. In such an interferometer, atoms are placed into a superposition of spatially separated atomic states, each of which has an associated quantum-mechanical phase term. When these states are brought back together at a later time, they will interfere with one another in a manner determined by the phase difference accrued between the states. A fountain geometry permits the atoms to remain in the superposition for a long time, allowing a large phase difference to evolve~\cite{Peters:Nature1999}. Such interferometers have been used to determine the Newtonian gravitational constant~\cite{Fixler:Science2007, Rosi:Nature2014}, measure the gravitational red shift~\cite{Mueller:Nature2010}, test quantum superposition over large distances~\cite{Kovachy:Nature2015}, and have been proposed to be used for testing general relativistic effects~\cite{Dimopoulos:PRL2007} and detecting gravitational waves~\cite{Tino:GRG2011}.

A molecular fountain will allow stringent tests of fundamental physics theories~\cite{Bethlem:EPJST2008, Tarbutt:NJP2013}. The sensitivity of any experiment looking for a frequency shift due to a certain physical phenomenon depends both on the size of the shift, i.e. the inherent sensitivity of the atom or molecule to a certain effect, and on the ability to measure this shift. Molecules' complex internal structure can make them very sensitive to aspects of fundamental physics~\cite{Steimle:JMS2014, DeMille:PhysToday2015}. For instance, in certain heavy molecules such as YbF and ThO the energy shift due to a permanent electric dipole moment of the electron is thousands of times larger than in heavy atoms~\cite{Hudson:Nature2011, Baron:Science2014}. Molecules are also used in the study of weak interactions leading to an anapole moment of the nucleus~\cite{DeMille:PRL2008}, in the search for a difference in transition frequency between chiral molecules that are each other's mirror-image~\cite{Quack:CPL1986, Daussy:PRL1999}, and for constraining a possible variation of the proton-to-electron mass ratio~\cite{Shelkovnikov:PRL2008, Jansen:JCP2014}.  The long interrogation times permitted by a molecular fountain will greatly enhance the sensitivity of these tests. A molecular fountain may also be used for interferometric measurements with molecules in free fall, for instance to test Einstein's Equivalence Principle for rotating objects~\cite{Hu:ChinPhysLett2012}. 

\begin{figure*}[bth!]
\centering
 \includegraphics[width=17cm]{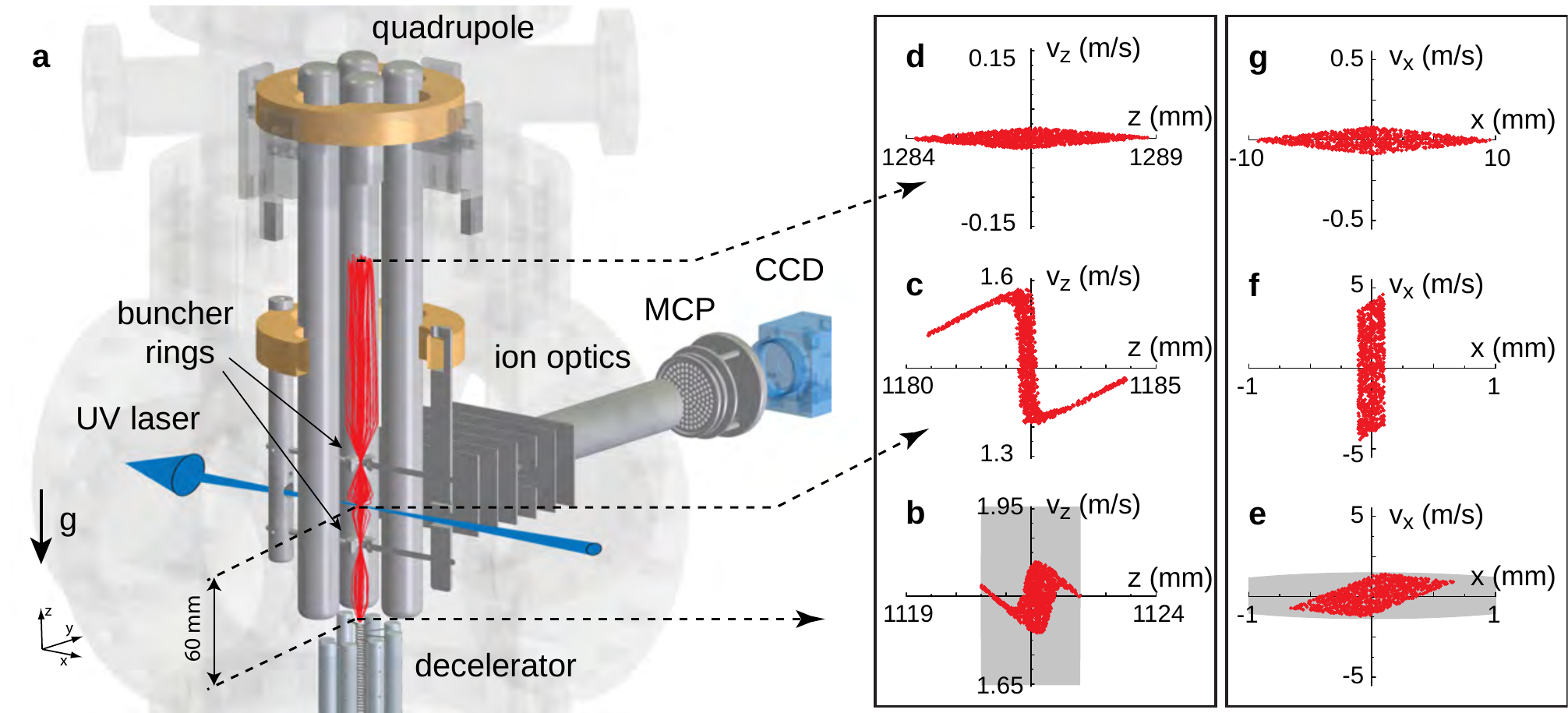}
 \caption{\label{fig1-setup}
Experimental set-up with simulated trajectories. a. Schematic view of the top part of the vertical beam machine showing the end of the traveling wave decelerator and the quadrupole lens system. The quadrupole lens consists of 4 cylindrical rods suspended by 2 ceramic discs. Two ring electrodes focus molecules in the $z$-direction. For a view on the inside, part of the quadrupole and the buncher has been cut. Molecules are ionized by a UV laser and imaged on a phosphor screen located behind a multi channel plate (MCP). The image is recorded using a charge coupled device (CCD) camera and a photo-multiplier tube (not shown). The red curves show a simulation of trajectories through the lens system for a beam launched with a velocity of 1.8\,m/s. b-g Phase-space plots showing the acceptance of the setup in both the longitudinal (b-d) and transverse directions (e-g), at three different heights. Note that the axes of panel g are scaled by a factor of 10 compared to panel e and f. The grey ellipses show the distribution of the packet of molecules at the exit of the decelerator.}
 \end{figure*}

Unfortunately, the enhanced sensitivity of molecules comes at a price: the rotational and vibrational degrees of freedom make molecules more difficult to cool~\cite{Carr:NJP2009, vandeMeerakker:NatPhys2008}. Laser cooling, while spectacularly successful at cooling atoms, is much less efficient for molecules~\cite{Shuman:Nature2010}. 

Here, we control the motion of molecules by exploiting the property that polar molecules experience a force in an inhomogeneous electric field~\cite{vandeMeerakker:NatPhys2008, Bethlem:PRL1999}. A supersonic beam of ammonia molecules is decelerated to rest using a combination of a conventional Stark decelerator,  consisting of a series of 100\,electrode pairs to which voltages of $\pm$10\,kV are applied, and a traveling wave decelerator, consisting of a series of 336\,rings to which oscillating voltages of $\pm$5\,kV are applied. For details of the molecular beam decelerator see Refs.~\cite{Quintero-Perez:PRL2013,Jansen:PRA2013}. Panel a of Fig.~\ref{fig1-setup} depicts the top part of the vertical molecular beam machine, showing the end of the traveling wave decelerator and the quadrupole focusing system. In the experiments presented here, ammonia ($^{14}$NH$_{3}$) molecules in the low-field seeking component of the $J$=1,$K$=1 state are brought to a standstill and trapped inside the traveling wave decelerator, about 10\,mm before the end. Once trapped, the molecules are adiabatically cooled to below 1\,mK by slowly ($\sim$2\,milliseconds) lowering the voltages from 5 to 1\,kV. Subsequently, the molecules are launched upwards with a variable acceleration to create a beam with a velocity in the range from 1.4 to 1.9\,m/s. Molecules with speed in this range will fly up 60-180\,mm before falling back under gravity.

\begin{figure*}[bth!]
\centering
 \includegraphics[width=13.5cm]{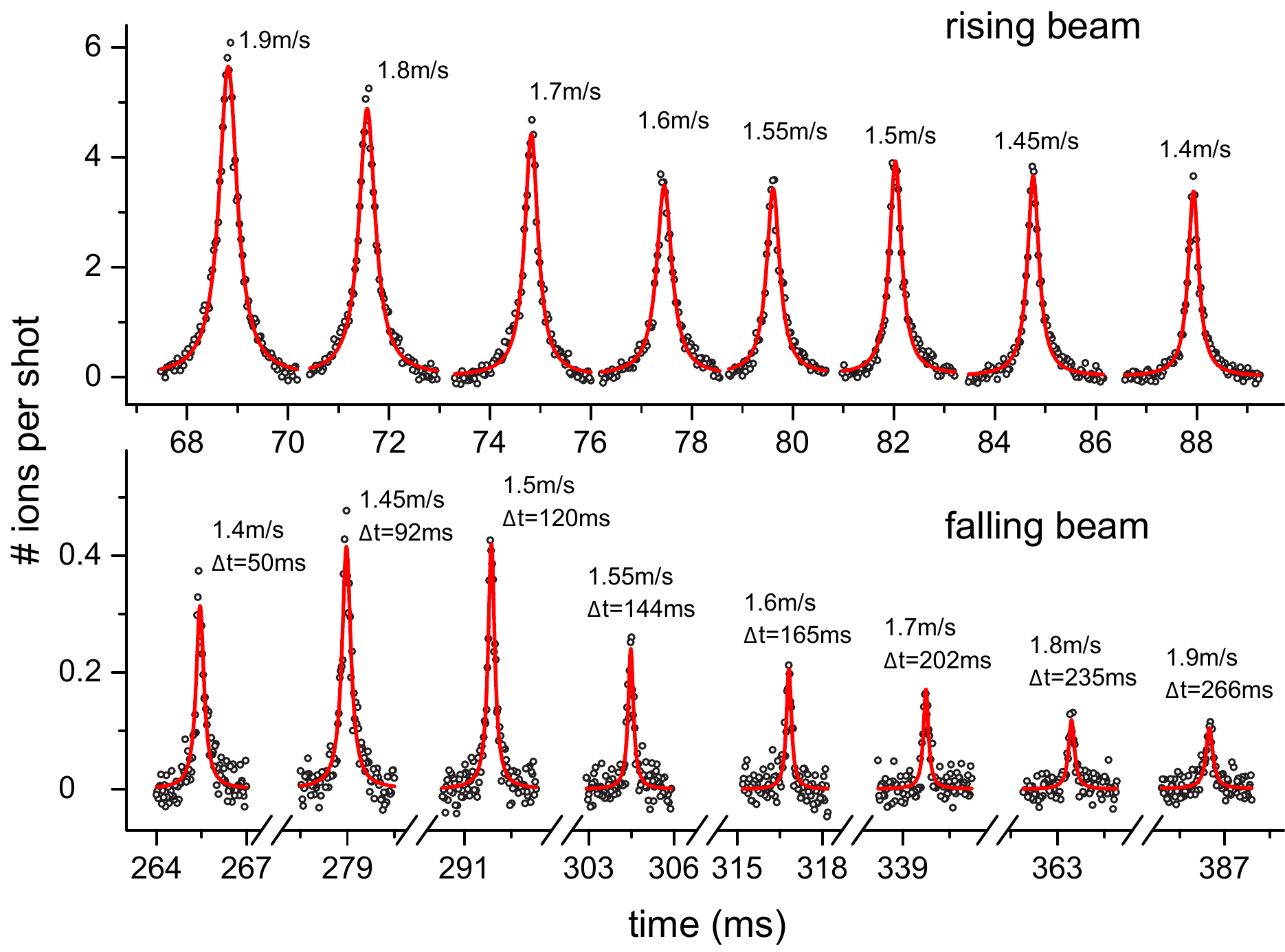}
 \caption{\label{fig2-TOFs}
Time-of-flight profiles of the rising and falling beams. Number of detected ammonia ions per shot as a function of the delay between opening the valve and firing the pulsed laser for the rising (top panel) and falling (lower panel) beams. Each data point (open circles) is averaged for 400 (1000) shots for the rising (falling) beam. The red lines show fits to Lorentz-profiles. Background signals of typically 0.7\,ions per shot (rising beam) and 0.15\,ions per shot (falling beam) have been subtracted.}
 \end{figure*}

The slow beam that exits the decelerator is focused using a combination of a linear quadrupole lens and two bunchers. The quadrupole lens consists of four 20\,mm diameter cylindrical rods which are spaced equidistantly on the outside of a 20\,mm diameter circle. The lens is switched on by applying voltages of $\pm$5\,kV to adjacent rods, resulting in an electric field that increases linearly away from the molecular beam axis. In electric fields below 10\,kV/cm, the Stark shift of $^{14}$NH$_{3}$ is almost perfectly quadratic and the molecules experience a harmonic potential that results in a transverse oscillation frequency of about 83\,Hz. In the longitudinal direction, the molecules are focused by applying a voltage of 4\,kV to two ring electrodes (6\,mm aperture) when they pass through each one, resulting in a longitudinal oscillation frequency of about 85\,Hz. The time-sequence is generated by a computer code that simulates the trajectories though the lens system and whose parameters are optimized using the molecular signal. The largest signal is obtained by focusing the beam at the buncher rings and the detection zone, and collimating the beam (i.e., minimizing the velocity spread) above the second buncher ring using a series of nine voltage pulses of varying duration. 

The red curves in panel a of Fig.~\ref{fig1-setup} show a simulation of trajectories through the lens system for a beam launched with a velocity of 1.8\,m/s. Molecules with this velocity will fly up ~165\,mm before falling back under gravity. They are in free fall in the top 67\,mm of their trajectory, i.e., no voltages are applied to the focusing elements during the 235\,milliseconds they need to traverse this path. Panels b-g show the corresponding distributions in phase space at three positions in the fountain: at the exit of the traveling wave decelerator, at the detection point, and at the apex. The origin of the $z$-axis is located at the position of the skimmer. The red dots show molecules that have stable trajectories throughout the lens system, and the grey ellipses show the distribution of the entire packet as it leaves the decelerator.  Note that the phase-space distribution at the detector is the same for the rising and falling beams (with the velocity in the $z$-direction having the opposite sign). The figure illustrates the extreme phase-space gymnastics that takes place in the setup: the transverse position spread at the apex is 50 times larger than at the detection zone. Consequently, the velocity spread is 50 times smaller, corresponding to a temperature of $<$10\,$\mu$K. In the longitudinal direction the velocity spread at the apex is about 8 times smaller than at the detection point,  corresponding to a temperature of $<$1\,$\mu$K. From the simulations, the acceptance of the lens system at 1.8\,m/s is found to be 0.05\,(mm$\cdot$m/s)$^3$. 

At the detection point, ammonia molecules in the $J$=1,$K$=1 state are ionized via a (2+1) \,REMPI scheme using pulsed UV laser radiation with wavelength 322\,nm. The resulting ions are focused onto a 2D detector using a series of ion lenses. By applying appropriate voltages to the ion optics, ions with the same initial velocity but a different position are focused at the same position on the detector (a technique known as velocity map imaging~\cite{Eppink:RSI1997}). A mask on the back side of the phosphor screen is used to transmit light from only the central part of the image onto a photomultiplier tube, thereby collecting light originating from molecules with a small transverse velocity. In this way we can discriminate signal from the rising and falling beam from signal originating from thermal gas in the chamber~\cite{QuinteroPerez:PCCP2012}. A CCD camera is used to optimize the position of the mask.
 
Fig.~\ref{fig2-TOFs} shows the ion signal as a function of time (open circles), catching the packets of molecules going up (upper panel) and coming down (lower panel). The shapes of the measured time-of-flight (TOF) profiles are determined by the focusing elements and are well described by Lorentzian fits (red curves). The velocity of the launched molecules and the time they are in free flight ($\Delta$t) are indicated above the TOF profiles. When the velocity of the beam is increased, the signal from the rising beam also increases as the divergence of the beam decreases. The signal from the falling beam, however, decreases as the molecules spend more time in free fall and spread out more before being recaptured. The signal from the falling beam is maximally 10\% of the signal from the rising beam with a launch velocity of 1.5\,m/s and drops to below 2\% with a launch velocity of 1.9\,m/s. This behavior is well reproduced by trajectory simulations when an exponential decay of the number of detected molecules with a $1/e$ decay time of 350\,milliseconds is included to account for losses due to collisions with background gas, non-linearity of the focusing forces and Majorana transitions. 

In conclusion, we have demonstrated a molecular fountain based on a Stark-decelerated molecular beam which enables the study of molecules in free fall for up to 266\,milliseconds. A Ramsey-type measurement during such a time interval would yield a linewidth of about 3\,Hz. With the obtained count rate of 0.1\,ions per cycle (one cycle takes 333\,milliseconds) a measurement time of 400\,seconds would suffice to determine the line-centre to within 1\,Hz. This offers the opportunity to measure inversion, rotational and vibrational transitions in ammonia with unprecedented accuracy. As these transitions have different dependences on the proton-to-electron mass ratio $\mu$, laboratory measurements can be used to set a stringent limit on a possible time-variance of $\mu$ due to cosmological expansion~\cite{Jansen:JCP2014} or as a result of dark matter~\cite{Derevianko:NatPhys2014}. Our method relies on adiabatic cooling and compression of a slow molecular beam using electrostatic lenses and can be applied to any polar molecule that is available at sufficiently high phase-space density, including heavy molecules like SrF~\cite{vandenBerg:JMS2014}, BaF and YbF~\cite{Bulleid:PRA2012}, which are used in low-energy tests of particle physics models.

\begin{acknowledgments}
The molecular fountain demonstrated here took many years to develop. We thank everybody who has contributed, in particular, Rob Kortekaas, Congsen Meng, Steven Hoekstra, Leo Huisman, Imko Smid, Jacques Bouma, Joost Buijs, Ruud van Putten, Ruth Buning, Masatoshi Kajita, Boris Sartakov, Andr\'{e} van Roij, Heinz Junkes, Georg Heyne, Henrik Haak and Gerard Meijer. This research has been supported by NWO via a VIDI Grant, by the ERC via a Starting Grant and by the Netherlands Foundation for Fundamental Research of Matter (FOM) (project 12PR2972 and program ``Broken mirrors and drifting constants'').
\end{acknowledgments}

\end{document}